\documentclass[english,prb, twocolumn, showpacs]{revtex4}
\usepackage[T1]{fontenc}
\usepackage[latin9]{inputenc}
\usepackage{verbatim}
\usepackage{float}
\usepackage{amsmath}
\usepackage{amssymb}
\usepackage{graphicx}
\usepackage{esint}

\makeatletter
\@ifundefined{textcolor}{}
{%
 \definecolor{BLACK}{gray}{0}
 \definecolor{WHITE}{gray}{1}
 \definecolor{RED}{rgb}{1,0,0}
 \definecolor{GREEN}{rgb}{0,1,0}
 \definecolor{BLUE}{rgb}{0,0,1}
 \definecolor{CYAN}{cmyk}{1,0,0,0}
 \definecolor{MAGENTA}{cmyk}{0,1,0,0}
 \definecolor{YELLOW}{cmyk}{0,0,1,0}
}

\usepackage{hyperref}

\makeatother

\usepackage{babel}
\begin{document}

\title{Proximity-induced superconductivity and Josephson critical current
in quantum spin Hall systems}

\author{Hoi-Yin Hui}

\affiliation{Condensed Matter Theory Center and Joint Quantum Institute, Department
of Physics, University of Maryland, College Park, Maryland 20742-4111,
USA.}

\author{Alejandro M. Lobos}

\affiliation{Condensed Matter Theory Center and Joint Quantum Institute, Department
of Physics, University of Maryland, College Park, Maryland 20742-4111,
USA.}

\author{Jay D. Sau}

\affiliation{Condensed Matter Theory Center and Joint Quantum Institute, Department
of Physics, University of Maryland, College Park, Maryland 20742-4111,
USA.}

\author{S. Das Sarma}

\affiliation{Condensed Matter Theory Center and Joint Quantum Institute, Department
of Physics, University of Maryland, College Park, Maryland 20742-4111,
USA.}

\date{\today}
\begin{abstract}
We consider recent experiments on wide superconductor-quantum spin
Hall insulator (QSHI)-superconductor Josephson junctions, which have
shown preliminary evidence of proximity-induced superconductivity
at the edge-modes of the QSHI system based on an approximate analysis
of the observed Fraunhofer spectra of the Josephson critical current
as a function of the applied magnetic field. Using a completely independent
exact numerical method involving a non-linear constrained numerical
optimization, we calculate the supercurrent profiles, comparing our
results quantitatively with the experimental Fraunhofer patterns in
both HgCdTe and InAs-GaSb based QSHI Josephson junctions. Our results
show good qualitative agreement with the experiments, verifying that
the current distribution in the 2D sample indeed has peaks at the
sample edges when the system is in the QSHI phase, thus supporting
the interpretation that superconductivity has indeed been induced
in the QSHI edge-modes. On the other hand, our numerical work clearly
demonstrates that it will be very difficult, if not impossible, to
obtain detailed quantitative information about the super-current distribution
just from the analysis of the Josephson Fraunhofer spectra, and, therefore,
conclusions regarding the precise width of the edge modes or their
topological nature are most likely premature at this stage.
\end{abstract}
\maketitle

\section{\label{sec:intro}Introduction}

Topological superconductors (TSC) are a special class of materials
characterized by the presence of a superconducting gap in the bulk,
and topologically protected gaples Majorana-fermion modes at the edges.\cite{Qi11_Review_TI_and_TSC}
In particular, Majorana zero-energy modes (MZMs) are predicted to
emerge as localized domain-wall-type excitations at the boundaries
of a one-dimensional (1D) TSC wire\cite{kitaev2001}, or at the vortex
cores of a two-dimensional (2D) TSC system\cite{volovik_JETP'99,Read00_Topological_SC_in_2D}.
MZMs have become the focus of an intense research activity due to
their non-Abelian braiding statistics, a property that could pave
the way to implement decoherence-free topological quantum computation
platforms.\cite{kitaev2001,Nayak08_RMP_Topological_quantum_computation,Wilczek'09,Ivanov_PRL'01}
While intrinsic TSC materials (with the appropriate topological order
parameters occurring naturally in the system) are scarce in nature,
in recent years this field has become increasingly active due to a
number of theoretical proposals predicting topological superconductivity
in hybrid structures such as the interface of topological insulator
(TI)/SC\cite{Fu08_Proximity-effect_and_MF_at_the_surface_of_TIs,Fu_PRB2009}
or semiconductor/SC\cite{Sau10_Proposal_for_MF_in_semiconductor_heterojunction,Lutchyn'10,Oreg'10,Sau10_long}
heterostructures. In these hybrid systems, an effective topological
superconductivity and MZMs arise from specific band structure engineering
which combines spin-orbit coupling, spin splitting, and proximity
effect induced by ordinary s-wave superconductors to produce the appropriate
topological order parameter. Experiments implementing some of these
proposals have reported preliminary evidence of MZMs in hybrid structures,
generating a great deal of interest in the community.\cite{Mourik12_Signatures_of_MF,Das12_Evidence_of_MFs,Deng12_ZBP_in_Majorana_NW,Finck13_ZBP_in_hybrid_NW_SC_device,Churchill2013,Rokhinson2012} 

In a seminal work\cite{Fu08_Proximity-effect_and_MF_at_the_surface_of_TIs},
Fu and Kane proposed to realize a TSC at the interface between a standard
s-wave SC and a TI. %
{} In the case of a quantum spin Hall insulator (QSHI), i.e., a 2D time-reversal
invariant TI with 1D helical edge modes, a 1D TSC with localized MZMs
is predicted to emerge at the interface with a proximate bulk SC (if
in addition the time reversal invariance is also broken somehow --
a necessary condition for the existence of localized MZMs). %
Experimental progress along this direction has been achieved recently
in HgTe/HgCdTe quantum wells\cite{Hart13_Induced_SC_at_QSH_edge}
and in InAs/GaSb quantum wells\cite{Pribiag14_Edge_mode_SC_in_2DTI},
two systems which have been predicted to enter the QSHI regime by
tuning the chemical potential in the band minigap or controlling the
well thickness. In the absence of a proximate SC, these 2D systems
exhibit signatures of edge-current transport in both electrical measurements\cite{Konig2007,Knez11_Evidence_of_edge_modes_in_InAsGaSb_QSHI}
and scanning SQUID microscopy experiments \cite{Nowack13_Edge_modes_in_QSHI_by_scanning_SQUID,Spanton14_Edge_modes_in_QSHI_by_scanning_SQUID}.
In particular for InAs/GaSb quantum wells contacted with superconducting
indium (In) leads, Knez \textit{et al}\cite{Knez12_Andreev_reflection_of_helical_edge_modes_in_QSHI}
demonstrated perfect Andreev reflection at the interface, as expected
for time-reversal protected QSH edge-modes where backscattering is
inhibited. This already indicates the exciting possibility of achieving
topological superconductivity in QSHI systems.

However, the above experiments do not conclusively show the presence
of supercurrents flowing only at the edges of the sample, as required
in the Fu-Kane scenario for the existence of topological superconductivity
and Majorana modes. Motivated by this question two experimental groups
\cite{Hart13_Induced_SC_at_QSH_edge,Pribiag14_Edge_mode_SC_in_2DTI}
recently performed measurements of the critical current $I_{c}\left(\Phi\right)$
flowing across SC-QSHI-SC Josephson junctions threaded by a magnetic
flux $\Phi=B.S$ (with $S$ the effective area of the junction and
$B$ an externally applied magnetic field perpendicular to the junction
area). The resulting Josephson critical current variation as a function
of the applied magnetic field, often referred to as the Fraunhofer
interference pattern $I_{c}\left(\Phi\right)$ vs $\Phi$ in the superconducting
electronics literature, can in principle be deconvoluted to extract
the current density profile $J\left(x\right)$ flowing across the
width $W$ of the junction, following a procedure suggested by Dynes
and Fulton \cite{Dynes71_SC_density_distribution} a long time ago.
The experiments, as analyzed using the Dynes-Fulton procedure, clearly
show the presence of supercurrent localized at the edges of the sample
precisely when the chemical potential is tuned across the (expected)
QSHI regime, and a more uniformly distributed bulk supercurrent profile
in the non-QSHI phase. Based on a quantitative deconvolution of the
observed Fraunhofer patterns using the Dynes-Fulton procedure, superconducting
edge-modes of widths 180-408 nm in the HgTe/HgCdTe system\cite{Hart13_Induced_SC_at_QSH_edge}
and 260nm in InAs/GaSb system\cite{Pribiag14_Edge_mode_SC_in_2DTI}
were reported. These experimental results, putatively showing the
presence of superconducting edge modes in the QSHI phase, are \emph{indirectly}
consistent with the presence of 1D TSC predicted by the Fu-Kane scenario,
but no definitive evidence for either the Majorana edge modes or topological
superconductivity has emerged yet.

It should be noted that the experiments in Refs.~\onlinecite{Hart13_Induced_SC_at_QSH_edge, Pribiag14_Edge_mode_SC_in_2DTI}
have made use of a certain number of highly simplifying assumptions
(inherent to the Dynes-Fulton analysis) which could potentially compromise
the interpretation of the experimental data, namely: 1) the Josephson
supercurrent density flowing in the junction (i.e., current-phase
relation) has been assumed to have a simple sinusoidal form\cite{Tinkham},
i.e., 

\begin{align}
J_{s}\left(x\right) & =J\left(x\right)\sin\left(\beta x+\varphi\right),\label{eq:sine_josephson}
\end{align}
where $\beta\equiv2\pi L_{J}B/\Phi_{0}$, with $L_{J}$ the effective
length of the junction and $\Phi_{0}$ the quantum of flux, and $\varphi$
an arbitrary uniform superconducting phase difference across the junction.
Strictly speaking, this assumption (i.e. the sinusoidal form for the
Josephson current) is only justified in the short-junction limit,
when $L\ll\xi$, where $L$ is the physical length of the junction
and $\xi$ is the coherence length in the SC contacts, and $L\ll\lambda_{J}$,
where $\lambda_{J}$ is the Josephson penetration depth\cite{Tinkham,likharev79_weaklinks_review}.
Although these are reasonable assumptions in many situations, in Refs.
\onlinecite{Hart13_Induced_SC_at_QSH_edge, Pribiag14_Edge_mode_SC_in_2DTI}
the parameter $\xi$ is not really known in the QSHI-SC junctions.
We do expect the Josephson penetration depth to be rather large in
general (\textasciitilde{} 1mm or so), and given that any proximity-induced
superconducting gap in the 2D semiconductor is likely to be small,
the coherence length (going as the inverse of the induced gap) should
also be large in the systems of interest. Therefore, the sinusoidal
approximation of Eq. (1) for the Josephson current seems reasonable,
but perhaps not absolutely compelling because of the lack of information
on the microscopic parameters of the junction. Another important requirement
for the validity of (\ref{eq:sine_josephson}) is that the transparency
of the contact must be small (i.e., transmission parameter $D\ll1$)\cite{likharev79_weaklinks_review}.
In the case of InAs/GaSb system the transmission at the SC/QSHI interface
was estimated to be of order unity ( i.e., $D\simeq0.7$), a regime
where the functional form of Eq.~(\ref{eq:sine_josephson}) is no
longer valid.\cite{Knez12_Andreev_reflection_of_helical_edge_modes_in_QSHI}
2) Furthermore, it is assumed in the analyses of Refs.~\onlinecite{Hart13_Induced_SC_at_QSH_edge, Pribiag14_Edge_mode_SC_in_2DTI}
that the dimensions of the junction ($W$ and $L_{J}$) are fixed
by the geometry of the sample and are therefore constants throughout
the experiment for a given 2D sample. However, a smooth dependence
of the effective junction parameters (e.g. $W$ and $L_{J}$) on other
experimental variables (e.g., gate voltage, external magnetic field,
etc.) cannot be excluded, resulting in \textit{effective} $\tilde{W}$
and $\tilde{L}_{J}$ varying with experimental parameters (such as
the magnetic field and the gate voltage) in some unknown manner. 3)
Another complication not included in the simple assumption of Eq.
(1) is the possibility of non-uniformity in the current distribution
along the other dimension (i.e. the current flow direction or the
$y$-direction) in the 2D plane which could arise from disorder and
density inhomogeneity invariably present in the semiconductor material
forming the QSH system. This is particularly germane for 2D QSH systems
because of the poor 2D screening properties and the invariable presence
of long-range Coulomb disorder in semiconductors due to random charged
impurities in the environment. Thus, in principle, $J(x)$ could depend
on two spatial variables, thus becoming $J(x,y)$, because of nonuniformity
and inhomogeneity, which would make the problem extremely complex
from a theoretical perspective. 4) Finally, even assuming that Eq.
(\ref{eq:sine_josephson}) is applicable, we note that the determination
of the supercurrent profile $J(x)$ starting from the measured Fraunhofer
spectra and using the Dynes-Fulton method is not unique since it is
essentially equivalent to an inverse scattering problem where all
phase information about the original unknown function one is interested
in obtaining is lost in the observed pattern being deconvoluted. The
critical current $I_{c}\left(\Phi\right)\equiv I_{c}\left(\beta\right)$
(i.e., the maximal supercurrent obtained by maximizing with respect
to the phase parameter $\varphi$) can be expressed as 

\begin{align}
I_{c}\left(\beta\right) & =\max_{\varphi}\left[\int dx\; J\left(x\right)\sin\left(\beta x+\varphi\right)\right]=\left|\mathcal{I}\left(\beta\right)\right|,\label{eq:Ic}
\end{align}
where

\begin{align}
\mathcal{I}\left(\beta\right) & \equiv\int dx\; J\left(x\right)e^{i\beta x},\label{eq:Fourier_transform}
\end{align}
is the Fourier transform of $J\left(x\right)$. The lack of knowledge
of the complex phase $\mathcal{I}\left(\beta\right)=\left|\mathcal{I}\left(\beta\right)\right|e^{i\theta\left(\beta\right)}$
means that different choices of $J\left(x\right)$ could reproduce,
in principle, the same \emph{observed} $I_{c}\left(\beta\right)$.
The standard way to overcome this problem is to consider the even
and odd parts of the current profile, $J\left(x\right)=J_{\text{e}}\left(x\right)+J_{\text{o}}\left(x\right)$,
and to assume that $J_{\text{e}}\left(x\right)\gg J_{\text{o}}\left(x\right)$.
In that case $\mathcal{I}\left(\beta\right)\simeq\int dx\; J_{\text{e}}\left(x\right)\cos\left(\beta x\right)$
is purely real, and the information about the phase can be easily
reconstructed taking the alternate lobes of the Fraunhofer pattern
as positive and negative, eventually interpolating linearly through
the minima. Then, the heights of the minima give a semiquantitative
measure of the odd part $J_{\text{o}}\left(x\right)$.\cite{Dynes71_SC_density_distribution,Hart13_Induced_SC_at_QSH_edge,Pribiag14_Edge_mode_SC_in_2DTI}
This procedure allows to uniquely determine an \textit{approximate}
$J_{\text{DF}}\left(x\right)$ where the subscript ``DF'' explicitly
refers to the fact that the resultant unique form for $J(x)$ arises
from an approximate Dynes-Fulton analysis assuming an almost-symmetric
current distribution. However, while this DF assumption can be expected
to hold in ideally symmetric cases, in general the presence of disorder
and inhomogeneity break the reflection symmetry with respect to the
center of the junction rendering this almost-even DF approximation
unjustified. In particular, when $J\left(x\right)\neq J_{\text{e}}\left(x\right)$,
Zappe\cite{Zappe75_Non_uniqueness_in_current_distribution} showed
explicitly that many different choices of $J\left(x\right)$ have
the same $I_{c}\left(\beta\right)$, thus emphasizing a long time
ago already that the DF prescription for obtaining the unknown Josephson
current distribution from the observed Fraunhofer Josephson pattern
could be suspect.

In view of the considerable importance of these recent experimental
developments in the context of the possible existence of MZM-carrying
topological superconductivity, it becomes crucial to critically examine
the validity of the results obtained in Refs.~\onlinecite{Hart13_Induced_SC_at_QSH_edge, Pribiag14_Edge_mode_SC_in_2DTI}
 using the Dynes-Fulton procedure. In this article we determine the
current density profile $J\left(x\right)$ using the Fraunhofer data
reported in Refs. \onlinecite{Hart13_Induced_SC_at_QSH_edge, Pribiag14_Edge_mode_SC_in_2DTI},
using an exact numerical optimization method that allows us to eliminate
the constraints of the Dynes-Fulton analysis. Specifically: a) we
allow for a non-sinusoidal current-phase relation, allowing for non-vanishing
transparencies $0\leq D\leq1$, b) we do not explicitly restrict the
functional form of $J\left(x\right)$ to be $J\left(x\right)\simeq J_{\text{e}}\left(x\right)$,
thereby lifting the symmetry constraint inherent in the Dynes-Fulton
analysis and finally, c) we impose a constraint in order to look for
solutions that avoid unphysical currents flowing outside the sample,
an undesirable by-product of the Dynes-Fulton prescription which is
apparent in many of the extracted $J_{{\rm DF}}(x)$ in Refs. \onlinecite{Hart13_Induced_SC_at_QSH_edge, Pribiag14_Edge_mode_SC_in_2DTI}
 where the calculated current flow often spills way outside the physical
sample dimensions used in the analysis.

We emphasize that, although our theoretical analysis uses a direct
numerical procedure to obtain the current distribution starting from
the observed Fraunhofer pattern in the QSHI Josephson junctions lifting
the nonessential assumptions and the limitations of the Dynes-Fulton
procedure, we still make a number of essential approximations ourselves
(all of them are also inherent in the Dynes-Fulton procedure and therefore
in the experimental analyses carried out in Refs. \onlinecite{Hart13_Induced_SC_at_QSH_edge, Pribiag14_Edge_mode_SC_in_2DTI}).
We assume that the current distribution is uniform in the y-direction
so that the basic Josephson junction approximation applies and we
also make the short-junction approximation. These approximations can
only be relaxed in a microscopic approach involving a direct numerical
solution of the applicable Bogoliubov-De Gennes (BdG) equations for
the SC-QSHI, but such a detailed approach is way beyond the scope
of the current work and may not in any case be suitable for extracting
the current distribution from the observed Fraunhofer pattern. Our
main accomplishment in the current work is an exact numerical analysis
of the Fraunhofer patterns reported in Refs. \onlinecite{Hart13_Induced_SC_at_QSH_edge, Pribiag14_Edge_mode_SC_in_2DTI}
going beyond the narrow constraints of the Dynes-Fulton procedure
and using a full numerical optimization scheme in going from the Franuhofer
patterns to the current distributions while at the same time generalizing
the theory to incorporate the finite transparency at the SC-QSHI interface
neglected in the Dynes-Fulton theory.

The paper is organized as follows: in Sec. \ref{sec:methods} we give
details about our numerical method, Sec. \ref{sec:Results} is devoted
to the presentation of our main results and finally, in Sec. \ref{sec:Conclusions}
we give a summary and our main conclusions.

\section{Methods\label{sec:methods}}

We now assume that the junction is in the short-junction limit $L\ll\left\{ \xi,\ \lambda_{J}\right\} $,
and we consider a more general expression for the current-phase relation
due to Haberkorn \textit{et al.}, valid for an arbitrary transparency
of the tunnel junction at the interface\cite{Haberkorn78_Current-phase_relation_finite_transparency,Golubov04_Josephson_junctions}

\begin{align}
I_{s}\left(\varphi\right) & =\frac{\pi\Delta}{2eR_{N}}\frac{\sin\left(\varphi\right)}{\sqrt{1-D\sin^{2}\left(\varphi\right)}}\tanh\left[\frac{\Delta}{2T}\sqrt{1-D\sin^{2}\left(\varphi\right)}\right],\label{eq:Haberkorn}
\end{align}
where $\Delta$ is the superconducting gap, $R_{N}=\left[e^{2}k_{F}^{2}SD/\left(4\pi^{2}\hbar\right)\right]^{-1}$
is the resistance of a contact of area $S$, and $T$ is the temperature.
This formula interpolates between the standard expression $I_{s}\left(\varphi\right)\propto\sin\left(\varphi\right)$
for the Josephson current-phase relationship as in Eq.~(\ref{eq:sine_josephson}),
valid in the tunnel limit $D\rightarrow0$, and the Kulik-Omelyanchuk
formula, $I_{s}\left(\varphi\right)\propto\sin\left(\varphi/2\right)\tanh\left[\frac{\Delta}{2T}\cos\left(\varphi/2\right)\right],$
valid at perfect transmission $D=1$.\cite{Kulik77_Current-phase_relation_perfect_transparency}
Based on Eq. (\ref{eq:Haberkorn}), we now generalize Eq. (\ref{eq:sine_josephson})
to

\begin{align}
J_{s}\left(x\right) & \propto J\left(x\right)\frac{\sin\left(\beta x+\varphi\right)}{\sqrt{1-D\sin^{2}\left(\frac{\beta x+\varphi}{2}\right)}}\quad\left(\text{at }T=0\right).\label{eq:Js_Haberkorn}
\end{align}
The starting point of our analysis is Eq. \ref{eq:Js_Haberkorn},
and not Eq. \ref{eq:sine_josephson} as used in Refs. \onlinecite{Hart13_Induced_SC_at_QSH_edge}
and \onlinecite{Pribiag14_Edge_mode_SC_in_2DTI} (and in the original
Dynes-Fulton treatment \cite{Dynes71_SC_density_distribution}). Note
that due to the square root in Eq. (\ref{eq:Js_Haberkorn}), the maximization
procedure with respect to $\varphi$ in Eq.~(\ref{eq:Ic}) needed
to find $I_{c}$, is \textit{not equivalent} to taking the amplitude
of the Fourier transform of $J\left(x\right)$, Eq.~(\ref{eq:Fourier_transform}),
and the analytical expression of $I_{c}$ becomes very complicated
and is neither transparent nor useful. Therefore, we need to perform
the maximization over $\varphi$ explicitly, redefining the critical
current using Eq.~(\ref{eq:Js_Haberkorn}) as

\begin{align}
I_{c}\left(\beta,D\right) & =\max_{\varphi}\left[\int_{-W/2}^{W/2}dx\;\frac{J\left(x,D\right)\sin\left(\beta x+\varphi\right)}{\sqrt{1-D\sin^{2}\frac{\beta x+\varphi}{2}}}\right].\label{eq:Ic_Haberkorn}
\end{align}
In order to find the current density profile $J\left(x\right)$, we
cast the problem in the form of a minimization procedure and define
a (dimensionless) \textit{cost functional} $C\left[J;D\right]$

\begin{align}
C\left[J;D\right] & =\frac{1}{2\beta_{\text{max}}}\int_{-\beta_{\text{max}}}^{\beta_{\text{max}}}d\beta\;\left[\frac{I_{c}\left(\beta,D\right)-I_{0}\left(\beta\right)}{I_{0}\left(0\right)}\right]^{2},\label{eq:cost}
\end{align}
where $I_{0}\left(\beta\right)$ is the experimentally obtained Fraunhofer
pattern (which we extract from the published data in Refs. \onlinecite{Hart13_Induced_SC_at_QSH_edge, Pribiag14_Edge_mode_SC_in_2DTI}).
To treat the problem numerically, we have discretized the variables
$x,\beta$ and $\varphi$ on a grid. For a given value of the transparency
$D$, we compute $C\left[J;D\right]$ by first performing the integral
over $x$ for all values of $\beta$ and $\varphi$ in Eq.~(\ref{eq:Ic_Haberkorn}).
Then for each value of $\beta$ the integral is maximized over $\varphi$.
Finally the integral over $\beta$ in Eq.~(\ref{eq:cost}) is performed,
and the value of $C\left[J;D\right]$ obtained. The numerical method
is initialized with a random initial function $J_{0}\left(x_{i}\right)$,
and a constrained non-linear optimization algorithm (``active-set''
algorithm in Matlab) is used to search for the minimum value $C\left[J\left(x\right);D\right]$,
within a relative tolerance $\epsilon=10^{-6}$. Similar to the Dynes-Fulton
case where the phase information cannot be reconstructed, we note
that our numerical procedure also does not, as a matter of principle,
uniquely determine the current profile because of the phase destruction
problem. However, for the Fraunhofer patterns that we generate numerically
from some assumed current patterns, our algorithm finds a current
distribution (from the resultant Fraunhofer patterns) where the cost
function is minimized to a very low value $C\left[J\left(x\right);D\right]\lesssim10^{-3}$,
which is negligible compared to the typical error bars in critical
current measurements. Although the final current distribution does
not typically precisely match the starting current distribution, this
low residual error confirms the validity of our procedure for finding
reasonably accurate solutions for the current density within our approximation
scheme of assuming short junctions and uniform current distributions.

Throughout the current work, we have used the consistency check of
numerically obtaining the current distribution from the Fraunhofer
spectra and then recalculating the Fraunhofer pattern from the obtained
current distribution comparing the resultant pattern with the starting
pattern to ensure the accuracy and the validity of our analysis. We
emphasize that although the first step of this procedure, i.e. obtaining
the actual current distribution from a given Fraunhofer pattern for
the maximum Josephson current as a function of the magnetic field
is non-unique because of the loss of phase memory, the reverse step
of recalculating the Fraunhofer pattern from the obtained current
distribution (i.e. Eqs.~(\ref{eq:Ic}) or (\ref{eq:Ic_Haberkorn})
depending on whether $D=0$ or not) is perfectly well-defined and
unique. We believe that this second step, which has not been presented
in the analyses of Refs. \onlinecite{Hart13_Induced_SC_at_QSH_edge, Pribiag14_Edge_mode_SC_in_2DTI},
is crucial in ensuring the integrity of the procedure leading to the
calculation of the current distribution in the QSHI sample.

\section{Results\label{sec:Results}}

\begin{figure}[t]
\centering{}\includegraphics[bb=0bp 5bp 720bp 250bp,clip,width=1\columnwidth]{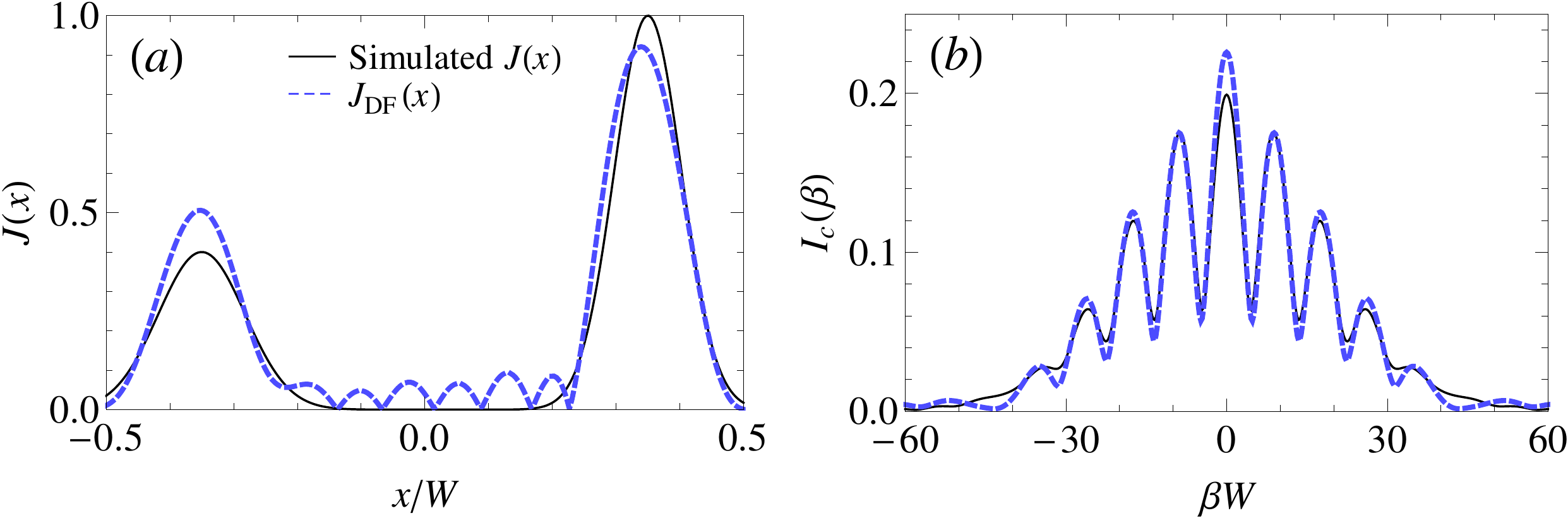}\caption{\label{fig:consistency}Consistency check of the Dynes-Fulton method.
(a) Simulated $J\left(x\right)$ (continuous black line) and reconstructed
$J_{\text{DF}}\left(x\right)$ (dashed blue line) density profiles.
(b) The corresponding Fraunhofer patterns. Although the original $J\left(x\right)$
and reconstructed $J_{\text{DF}}\left(x\right)$ profiles are qualitatively
similar, there is no quantitative agreement.}
\end{figure}
We first focus on the internal consistency of the Dynes-Fulton method
in situations which deviate from the ideal condition $J\left(x\right)=J_{\text{e}}\left(x\right)$.
Given the assumptions implicit in the method, our point is that the
resulting $J_{\text{DF}}\left(x\right)$ is not a stable self-consistent
solution upon iteration of the procedure. In other words, injecting
the approximate $J_{\text{DF}}\left(x\right)$, obtained with the
Dynes-Fulton method, back into Eq. (\ref{eq:Fourier_transform}) does
not allow to recover the original Fraunhofer pattern $I_{c}\left(\beta\right)$,
from which $J_{\text{DF}}\left(x\right)$ was first extracted. This
is an important point (i.e., a ``sanity'' check to see how serious
an issue the non-uniqueness aspect of the problem turns out to be)
in order to test the robustness of the method. To illustrate this
point, in Fig.~\ref{fig:consistency}(a) we show a simulated current
density profile in continuous blue line, and in Fig.~\ref{fig:consistency}(b),
the corresponding Fraunhofer pattern {[}obtained using Eq. (\ref{eq:Fourier_transform}){]}
is also shown as a continuous blue line. We now apply the Dynes-Fulton
method, and the ``reconstructed'' profile $J_{\text{DF}}\left(x\right)$
is shown in Fig. \ref{fig:consistency}(a) (dashed black line). Note
that the original $J\left(x\right)$ and reconstructed $J_{\text{DF}}\left(x\right)$
profiles are \textit{only qualitatively} similar, there is no \textit{quantitative}
agreement at all: the information about the fine details of the profile
(e.g., relative heights or widths of the peaks) is not consistently
reproduced. This is actually not surprising given that the initial
profile $J\left(x\right)$ does not fulfill the required reflection
symmetry condition $J\left(x\right)=J_{\text{e}}\left(x\right)$ for
Dynes-Fulton the method to be exact. Moreover, additional relevant
physical mechanisms such as finite transparency (not considered in
Fig.~\ref{fig:consistency}) can only worsen the situation. For completeness,
in Fig. \ref{fig:consistency}(b) we also show the Fraunhofer pattern
(dashed black line) recalculated from the extracted $J_{\text{DF}}\left(x\right)$.
Again, the original and final Fraunhofer patterns are qualitatively
very similar, but the fine quantitative details are not reproduced.
We believe that this comparison between the original and reconstructed
Fraunhofer patterns could serve as a useful tool to check the internal
consistency of the procedure in concrete experimental situations.
In the Appendix \ref{sec:SuppMat} we have performed such an internal
consistency comparison using a direct digitalization of the data reported
in the QSH-SC Josephson junction experimental papers\cite{Hart13_Induced_SC_at_QSH_edge,Pribiag14_Edge_mode_SC_in_2DTI}.

\begin{figure}[t]
\begin{centering}
\includegraphics[bb=20bp 0bp 710bp 450bp,clip,width=1\columnwidth]{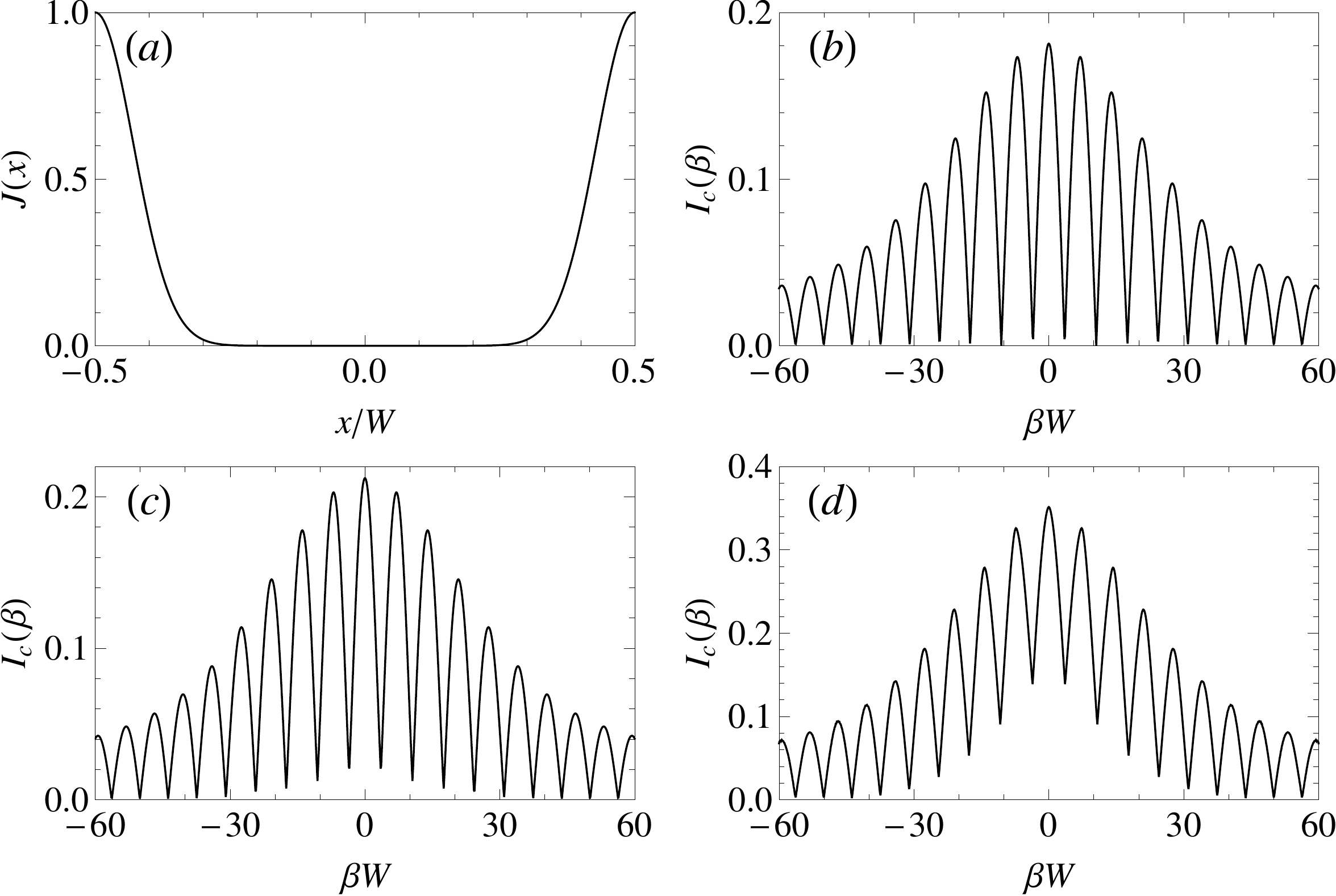}
\par\end{centering}

\caption{(a) The chosen current density. (b) the critical current against $\beta$
with $D=0$. (c) the critical current against $\beta$ with $D=0.5$.
(d) the critical current against $\beta$ with $D=1$. \label{fig:nodelift}}
\end{figure}

We now focus on the effects of a finite transparency $D$ on the Fraunhofer
patterns. As mentioned before, this is an important element of physics
missing from the earlier analyses in Refs. \onlinecite{Hart13_Induced_SC_at_QSH_edge, Pribiag14_Edge_mode_SC_in_2DTI}
since the experimental QSH-SC junctions may very well be closer to
the fully transparent ($D=1$) limit than the fully non-transparent
($D=0$) Dynes-Fulton limit. In Fig \ref{fig:nodelift}(a), we show
an arbitrarily chosen current distribution profile $J\left(x\right)$
with the current flowing at the edges of the sample. Here we have
chosen a purely even profile $J\left(x\right)=J_{\text{e}}\left(x\right)$
for the purpose of illustration. We next use Eq. (\ref{eq:Ic_Haberkorn})
to compute the corresponding Fraunhofer pattern for transparencies
$D=0$, $D=0.5$ and $D=1.0$, and plot them in Figs. \ref{fig:nodelift}(b),
\ref{fig:nodelift}(c), and \ref{fig:nodelift}(d), respectively.
Note that when $D\neq0$, these patterns exhibit a sizeable ``node-lifting''
effect. In the simplest Dynes-Fulton analysis (valid only for $D=0$),
the lifted nodes would have been \textit{mistakenly} attributed to
asymmetries in $J\left(x\right)$, i.e., non-vanishing odd component
$J_{\text{o}}\left(x\right)$. We point out that this node-lifting
mechanism has not been discussed previously in Refs. \onlinecite{Hart13_Induced_SC_at_QSH_edge, Pribiag14_Edge_mode_SC_in_2DTI},
and is completely different from the discussion on node-lifting in
Ref. \onlinecite{Lee14_Topological_node_lifting_in_Fraunhofer_pattern}.
With these ideas in mind, we now focus on the analysis of the experimental
Fraunhofer patterns. 

\begin{figure}[t]
\begin{centering}
\includegraphics[bb=0bp 0bp 500bp 390bp,clip,width=1\columnwidth]{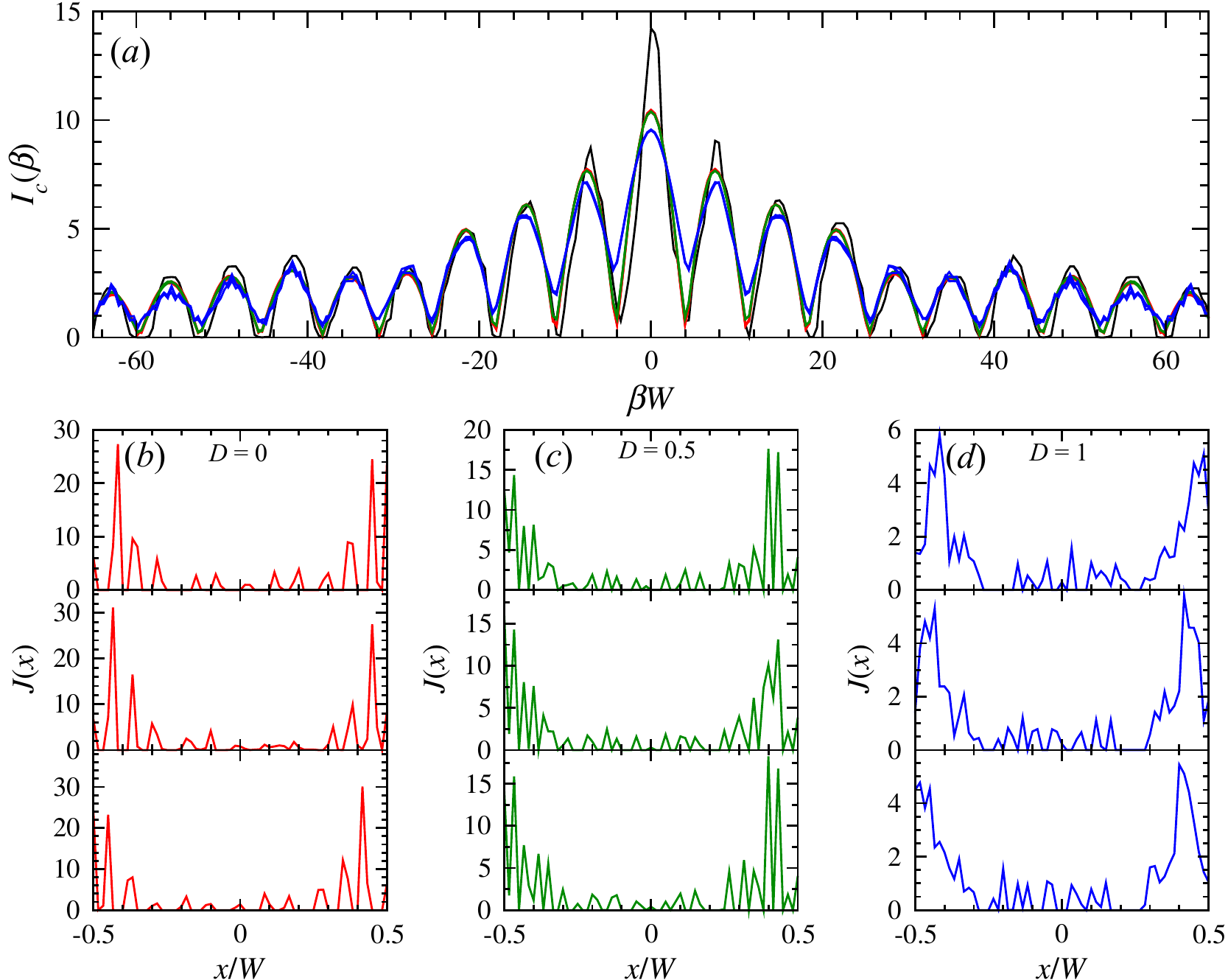}
\par\end{centering}

\caption{\label{fig:Yacoby_SQH}(a) Black line: Fraunhofer pattern obtained
from digitalization of the experimental results in Fig. 2(c) of Ref.
\onlinecite{Hart13_Induced_SC_at_QSH_edge}, where the sample is believed
to be in the SQH insulating regime. Red, green and blue lines correspond
to the Fraunhofer patterns obtained from the profiles below, found
with our numerical minimization, for $D=0$, $D=0.5$ and $D=1$,
respectively. In plots (b), (c) and (d) we show the corresponding
profiles for transparencies $D=0$, $D=0.5$ and $D=1.0$, respectively.
Slightly different profiles are obtained in each case (here we show
only three for each value of $D$) as a consequence of the abovementioned
non-uniqueness of the procedure.}
\end{figure}

We first analyze the HgCdTe-CdTe data by Hart \textit{et al}.\cite{Hart13_Induced_SC_at_QSH_edge}
in the regime where the system is in the QSHI phase. In Fig. \ref{fig:Yacoby_SQH}(a),
the black line corresponds to the experimental Fraunhofer pattern
shown in the figure 2(c) in Ref. \onlinecite{Hart13_Induced_SC_at_QSH_edge}.
In Figs. \ref{fig:Yacoby_SQH}(b), \ref{fig:Yacoby_SQH}(c), and \ref{fig:Yacoby_SQH}(d)
we show the profiles resulting from our minimization procedure assuming
transparencies $D=0$, $D=0.5$ and $D=1.0$, respectively. Slightly
different profiles are obtained for each transparency (here we show
only three for each value of $D$) as a consequence of the abovementioned
non-uniqueness of the procedure, but besides this fact, the overall
features of the different plots for each value of $D$ are consistent.
Qualitatively speaking, our results agree with those of Hart \textit{et
al}. in the sense that all the numerically-obtained current density
profiles \textit{indeed exhibit distinct peaks at the edges}. 
Considering that our numerical search is free of predetermined assumptions
inherent in the Dynes-Fulton prescription, we believe this qualitative
agreement to be an important confirmation of the experimental results
as definitively establishing the presence of edge super-current transport
in the HgCdTe-based SC-QSHI-SC Josephson junction system. We should
add for emphasis here that we have tried hard to obtain current distributions
without peaks at the sample edges in the QSHI phase, but have consistently
failed to find a single situation where our exact numerical procedure
(with only the short junction and uniform current constraints) produces
current distributions without peaks at the edges which are consistent
with the experimentally observed Fraunhofer patterns in the QSHI phase
reported in Ref. \onlinecite{Hart13_Induced_SC_at_QSH_edge}. Moreover,
our reconstructed Fraunhofer patterns {[}shown in Fig. \ref{fig:Yacoby_SQH}(a)
as red, green and blue lines{]} using the numerically extracted current
distributions compare fairly well with the experimental one (continuous
black line). Reconstructed Fraunhofer patterns with the same color
(obtained with the same value of $D$) fall on top of each other,
and cannot be distinguished at the scale of the plots shown in Fig.~\ref{fig:Yacoby_SQH}.
We take this agreement as an internal consistency check of our procedure.
Finally, we note that the better agreement of the experimental Fraunhofer
pattern with the red curve in Fig. \ref{fig:Yacoby_SQH}(a) (i.e.
$D=0$) might be indicative of a low transparency of the contacts
in the experiment of Ref. \onlinecite{Hart13_Induced_SC_at_QSH_edge}.

However, one should take these results with caution, in particular
when considering the fine quantitative details. For instance, we note
that the reconstructed Fraunhofer patterns in Fig. \ref{fig:Yacoby_SQH}(a)
do not correctly reproduce the behavior of the original experimental
pattern near $\beta\approx0$. We have checked that this is not a
problem of the numerical method {[}e.g., failure to converge to the
``true'' minimum in Eq. (\ref{eq:cost}){]}, because our code always
reproduces the patterns that are numerically generated from a wide
variety of current densities that are localized within the width of
the wire. Therefore, the discrepancy between the experimental data
and the reproduced Fraunhofer pattern suggests a break-down of (at
least) one of the two main assumptions (a) the gate voltage independence
of $L_{J}$ or $W$, or (b) the short junction current-phase relation
{[}Eq. (\ref{eq:sine_josephson}) or even the more generic Eq. (\ref{eq:Js_Haberkorn}){]}.
Our best current guess right now is that the sample size parameters
($L_{J}$ and/or $W$) do indeed vary with gate voltage (and magnetic
field) leading to this inconsistency, but more work is needed to clarify
this issue.

\begin{figure}[t]
\begin{centering}
\includegraphics[bb=0bp 0bp 555bp 450bp,clip,width=1\columnwidth]{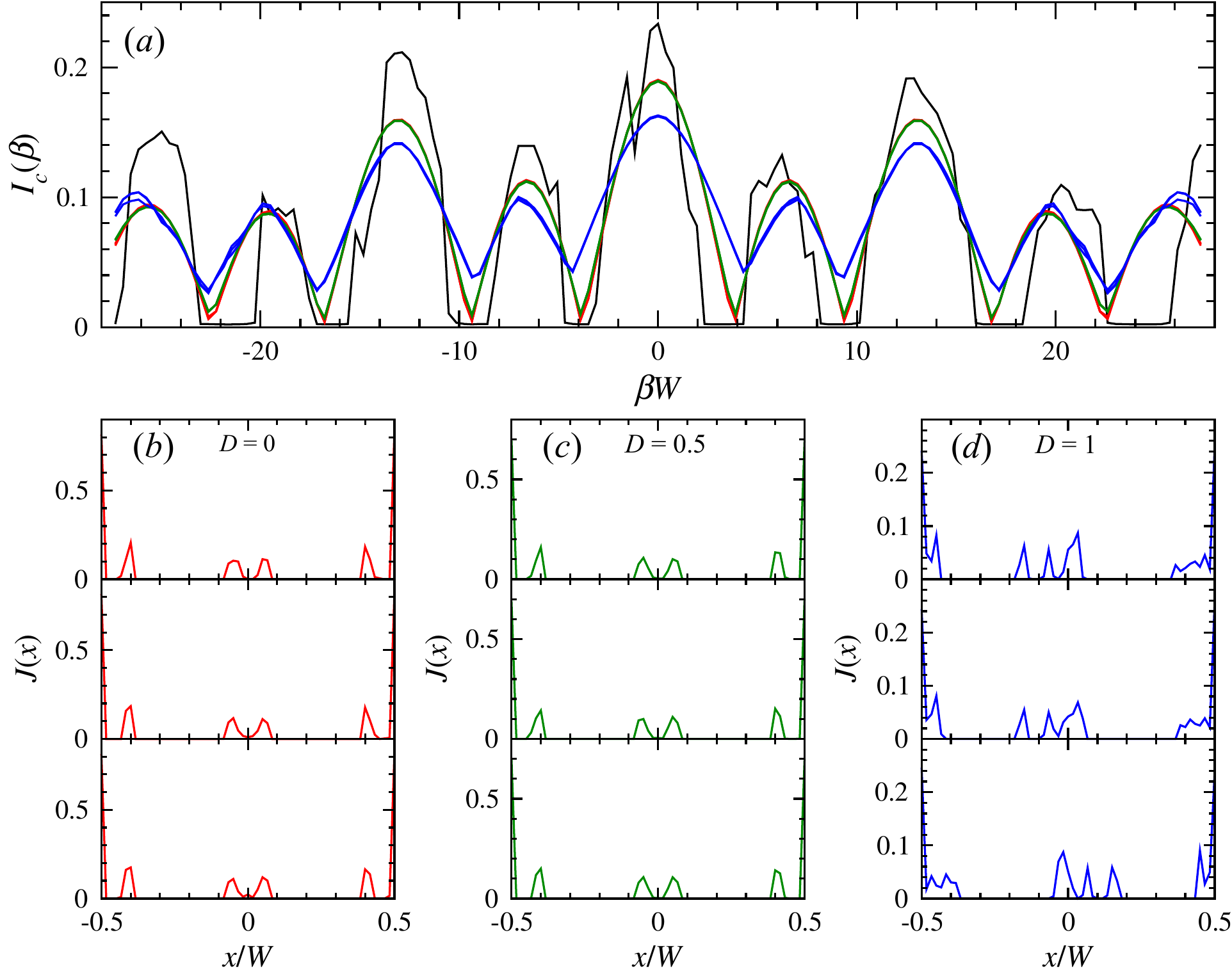}
\par\end{centering}

\caption{\label{fig:Kouwenhoven5b_SQHI}(a) Black line: Fraunhofer pattern
obtained from direct digitalization of the experimental results in
Fig. 5(a) of Ref. \onlinecite{Pribiag14_Edge_mode_SC_in_2DTI}, when
the sample is in the SQH insulating regime. Red, green and blue lines
correspond to the Fraunhofer patterns obtained from the profiles below,
found with our numerical minimization, for $D=0$, $D=0.5$ and $D=1$,
respectively. In plots (b), (c) and (d) we show the corresponding
profiles for transparencies $D=0$, $D=0.5$ and $D=1.0$, respectively.
Slightly different profiles are obtained in each case (here we show
only three for each value of $D$) as a consequence of the abovementioned
non-uniqueness of the procedure.}
\end{figure}
Our numerics also agree qualitatively (but less quantitatively compared
with Ref. \onlinecite{Hart13_Induced_SC_at_QSH_edge}) with the results
by Pribiag \textit{et al} for the InAs-GaSb system\cite{Pribiag14_Edge_mode_SC_in_2DTI}.
In particular in this case, we have focused on Fig. 5(a) in Ref. \onlinecite{Pribiag14_Edge_mode_SC_in_2DTI},
where the authors report an intruiguing even-odd effect in the Fraunhofer
pattern when the sample is in the extreme n-doped regime (which can
be thought of as the emergence of a $2\Phi_{0}-$periodicity in the
Josephson current). The authors attribute this behavior to two possible
physical mechanisms: a) a topologically-trivial mechanism corresponding
to a very particular spatial current distribution containing three
peaks, or b) the presence of fractional-Josephson effect, which is
indeed expected when MZMs emerge at the ends of the induced TSC in
the Fu-Kane scenario\cite{Fu08_Proximity-effect_and_MF_at_the_surface_of_TIs}.
Both scenarios seem unlikely as mentioned already in Ref. \onlinecite{Pribiag14_Edge_mode_SC_in_2DTI}.
In the case of the three-peak current distribution function, a high
degree of fine-tuning seems to be required to reproduce the Fraunhofer
pattern. The authors of Ref. \onlinecite{Pribiag14_Edge_mode_SC_in_2DTI}
argue that since the even-odd effect has been observed in at least
two samples, it seems improbable that these finely tuned parameters
occur in \textit{both} samples simultaneously. On the other hand,
the topological mechanism based on Josephson-coupled MZMs requires
a quasiparticle poisoning time-scale which exceeds the actual experimental
measurement time, an extremely unlikely scenario unless very special
efforts are undertaken to suppress quasiparticle poisoning deliberately.
Recently, starting from a microscopic description for transport across
a SC-TI-SC junction, Tkachov \textit{et al}\cite{Tkachov14_Edge_supercurrents_in_2DTI}
have predicted a temperature-driven crossover from $\Phi_{0}-$ to
$2\Phi_{0}-$periodic behavior in the Fraunhofer pattern, therefore
providing an alternative mechanism for this behavior. In their treatment,
the authors attibute this behavior to the well-known progressive skewness
of the current-phase relationship in \textit{long junctions} as the
temperature is lowered below $k_{B}T\ll\hbar v/\pi L$ \cite{Ishii70_SNS_junction,Svidzinsky73_SNS_junction},
fitting the experimental curves with parameters $L_{J}=600$ nm and
$\xi=80$ nm. We point out, however, that the prescription used by
Tkachov \textit{et al} involves fixing the phase difference $\varphi$
of the SCs by maximizing the supercurrent \textit{only at zero field},
and using that value of $\varphi$ for finite-$B$ calculations of
the supercurrent. Since the SC phase difference is a gauge dependent
quantity, it is not clear that such a prescription obeys gauge invariance.
In addition, we note that their theoretical model does not explain
the observed dependence on the gate voltage, and does not allow to
extract the current profile. Also, the long junction limit with very
short coherence length seems inconsistent with the analyses of Refs.
\onlinecite{Hart13_Induced_SC_at_QSH_edge, Pribiag14_Edge_mode_SC_in_2DTI}
as well as our own theory in the current paper. At this stage, therefore,
our preference (see below) for the observed even-odd effect in Ref.
\onlinecite{Pribiag14_Edge_mode_SC_in_2DTI} is that for some unknown
reason the spatial current distribution does indeed have a 3-peak
structure which is obviously not ruled out by any known mechanism.
More future work will be necessary to clarify this issue decisively,
but at this stage we believe that the short-junction approximation
is perhaps not a poor approximation for the QSHI systems under consideration
given the expected long coherence length and the long Josephson penetration
depth applying in the problem.

Our analysis of this case is shown in Fig. \ref{fig:Kouwenhoven5b_SQHI}.
In Fig. \ref{fig:Kouwenhoven5b_SQHI}(a), the black line corresponds
to the Fraunhofer pattern obtained from direct digitalization of the
experimental results\cite{Pribiag14_Edge_mode_SC_in_2DTI} in figure
5(a). Figs. \ref{fig:Kouwenhoven5b_SQHI}(b), \ref{fig:Kouwenhoven5b_SQHI}(c)
and \ref{fig:Kouwenhoven5b_SQHI}(d) correspond to different current
densities $J\left(x\right)$ obtained with our numerical method assuming
transparencies $D=0$, $D=0.5$ and $D=1.0$, respectively. In addition,
we also observe a qualitative agreement of our numerical results with
the experimental Fraunhofer pattern. However, we point out here the
striking apparent lack of internal consistency of the reported data,
which becomes evident when comparing the experimental Fraunhofer pattern
with the reconstructed one, for which we have no satisfactory explanation
(see Fig. \ref{fig:Kouwenhoven_consistency} in the Appendix). This
inconsistency points toward a possible inapplicability of at least
one of the theoretical constraints to the experimental situation applying
to Ref. \onlinecite{Pribiag14_Edge_mode_SC_in_2DTI}, and the possibility
that the system is in the long junction limit (or perhaps there is
considerable current nonuniformity) cannot be ruled out at this stage.

\section{Conclusions\label{sec:Conclusions}}

We have analyzed the validity and the consistency of the Dynes-Fulton
procedure\cite{Dynes71_SC_density_distribution} applied to recent
Fraunhofer pattern experiments on wide SC-QSHI-SC Josephson junctions\cite{Hart13_Induced_SC_at_QSH_edge,Pribiag14_Edge_mode_SC_in_2DTI},
which have shown evidence for proximity-induced superconductivity
at the edge-modes of the QSHI system. Using a completely independent
exact numerical method (i.e., a non-linear constrained numerical procedure,
see Section \ref{sec:methods}) to analyze the experimental data extracted
from Refs.~\onlinecite{Hart13_Induced_SC_at_QSH_edge, Pribiag14_Edge_mode_SC_in_2DTI},
we obtain the current profiles in Figs. \ref{fig:Yacoby_SQH} and
\ref{fig:Kouwenhoven5b_SQHI} which agree qualitatively (but, not
quantitatively) with the conclusions in Refs.~\onlinecite{Hart13_Induced_SC_at_QSH_edge, Pribiag14_Edge_mode_SC_in_2DTI}.

Generally speaking, our results \textit{qualitatively} reproduce current
distributions with peaks at the 2D sample edges when the system is
in the QSHI phase. Given that our method is completely free from the
nonessential approximations of the Dynes-Fulton analysis (i.e., vanishing
transparency $D=0$ of the SC-QSHI contacts, symmetric current distributions,
and profiles extending beyond the physical region $-W/2<x<W/2$), we find
that our results support the existence of proximity-induced superconductivity
at the edge-modes of the QSHI. This is of course highly encouraging
from the perspective of the eventual emergence of topological superconductivity
with Majorana zero modes in quantum spin Hall systems.

Our work also clearly shows that the Dynes-Fulton prescription for
obtaining the supercurrent spatial distribution from the observed
Fraunhofer pattern has a number of intrinsic limitations, stemming
from its simplicity and the various underlying approximations. For
instance, we have analyzed the internal consistency of the method
by comparing the original (i.e., experimental) Fraunhofer pattern
with the reconstructed pattern obtained from the extracted current
distribution $J_{\text{DF}}\left(x\right)$ (see Figs. \ref{fig:Yacoby_consistency}
and \ref{fig:Kouwenhoven_consistency} in the Appendix). While in
some cases the method seems to be reasonably successful in reproducing
the original pattern, we note that in general there is a lack of quantitative
agreement, which seems to be generic and inherent to the method itself
stemming from the non-unique nature of the inverse scattering problem
it is trying to solve. Moreover, we note that the quantitative prediction
of the Fraunhofer pattern analyses in Refs.~\onlinecite{Hart13_Induced_SC_at_QSH_edge, Pribiag14_Edge_mode_SC_in_2DTI},
such as the relative height or the width of the peaks, must also be
taken with a great deal of caution. We note that the reported widths
of the edge-modes ($\sim260$ nm in the case of InAs/GaSb \cite{Pribiag14_Edge_mode_SC_in_2DTI}
and of $\sim180-408$ nm in the case of HgTe/HgCdTe \cite{Hart13_Induced_SC_at_QSH_edge})
are about one order of magnitude larger than the expected QSH edge
width obtained from band structure considerations\cite{Zhou08_Finite_size_effects_on_edge_states_in_QSH_system}.
This discrepancy remains unexplained at the moment, and cannot be
addressed within the present analysis. Among the possible physical
mechanisms which could account for wider edge modes, we mention strong
band-bending effects at the edge (which may induce multichannel transport)
and disorder/puddle effects (which make the edge modes go around the
puddles, effectively enhancing the width). At this stage, the theory
is not capable of providing a quantitatively accurate prediction for
the size of the super-current distribution at the QSHI edges in the
presence of band-bending, disorder, and proximity effect. One critically
disturbing aspect of the very wide effective edge size (an order of
magnitude larger than the theoretically predicted value for the topological
helical edge mode which should be of the order of the QSHI ``coherence
length'' defined by $v/E_{g}$, where $v$ and $E_{g}$ are the edge
velocity and the bulk gap respectively), deduced from the Fraunhofer
experiments of Refs. \onlinecite{Hart13_Induced_SC_at_QSH_edge} and
\onlinecite{Pribiag14_Edge_mode_SC_in_2DTI}, is that the simple and
widely used model of a protected helical ``topological'' edge mode
obviously would not apply to the real samples implying (among many
things) that the whole issue of inducing topological edge mode superconductivity
with the associated localized Majorana zero end modes needs to be
carefully thought through\cite{Hui14_Disorder_in_TI_SC_hybrids} taking
into account the theoretically-unexplained very large widths of these
edge modes for both HgTe-HgCdTe and InAs-GaSb QSHI systems. More quantitatively
accurate experiments determining the QSHI edge electronic structure
would be highly desirable in this context because it is unclear at
this stage how accurate the estimated edge widths based on Fraunhofer
measurements really are because of the many approximations in the
theory as well as the non-uniqueness of the theory extracting a spatial
supercurrent distribution from the measured Fraunhofer pattern as
discussed in details in our work. It should perhaps be mentioned here
as a cautionary note that for the corresponding quantum Hall (QH)
edge modes, there has not been any unifying picture connecting theory
and experiment for the electronic structure of the QH edges in spite
of 25 years of intense activity (see, for example, H. Choi \textit{et
al}\cite{Choi14_Robust_electron_pairing_in_IQH_regime} and references
therein).

To summarize: Our theoretical conclusion is more or less consistent
with those in Refs.~\onlinecite{Hart13_Induced_SC_at_QSH_edge, Pribiag14_Edge_mode_SC_in_2DTI} showing definitive evidence for the existence of super-current edge
modes in the QSHI phase of HgCdTe-CdTe and InAs-GaSb quantum well
systems in spite of considerable non-uniqueness associated with the
extraction of the current distribution pattern from the observed Fraunhofer
pattern. In addition, we have explicitly pointed out the inadequacy
and inconsistency of the simplistic Dynes-Fulton procedure for analyzing
the Fraunhofer spectra in wide Josephson junctions.
\begin{acknowledgments}
The authors acknowledge support from Microsoft Q, NSF-JQI-PFC, and
NSA-LPS-CMTC. J. D. Sau acknowledges Anton Akhmerov for valuable discussions.
\end{acknowledgments}
\appendix

\section{Consistency check of the experimentally reported data\label{sec:SuppMat}}

\begin{figure}[H]
\includegraphics[width=0.95\columnwidth]{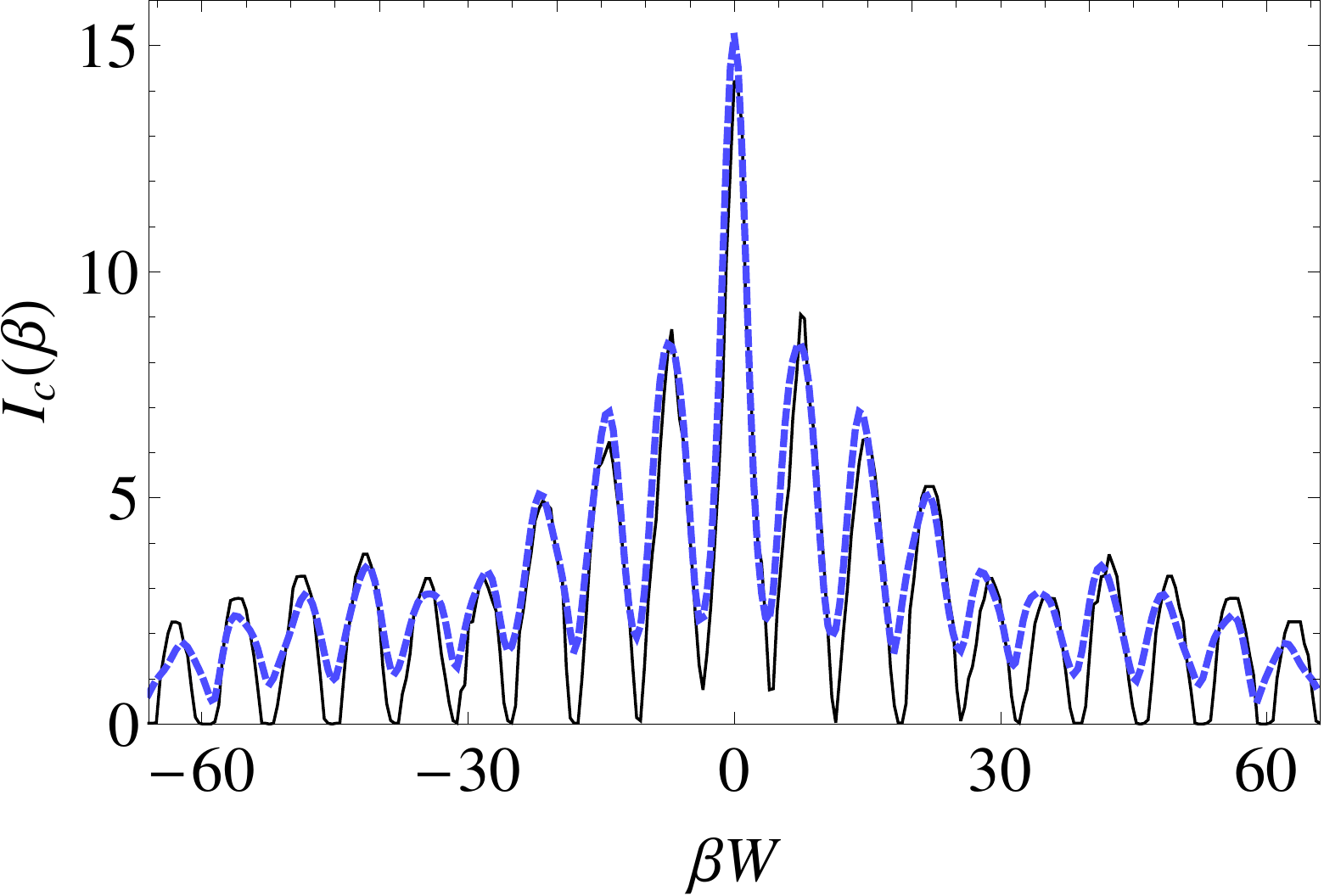}

\caption{\label{fig:Yacoby_consistency}Fraunhofer pattern obtained from digitalization
of the experimental results in Fig. 2(c) of Ref. \onlinecite{Hart13_Induced_SC_at_QSH_edge}
(shown as a black line), where the sample is believed to be in the
SQH insulating regime. The blue dashed line corresponds to the reconstructed
Fraunhofer pattern obtained from the reported current profile $J_{\text{DF}}\left(x\right)$
in figure 2.d in that reference.}
\end{figure}

\begin{figure}[H]
\includegraphics[bb=0bp 0bp 450bp 350bp,clip,width=0.95\columnwidth]{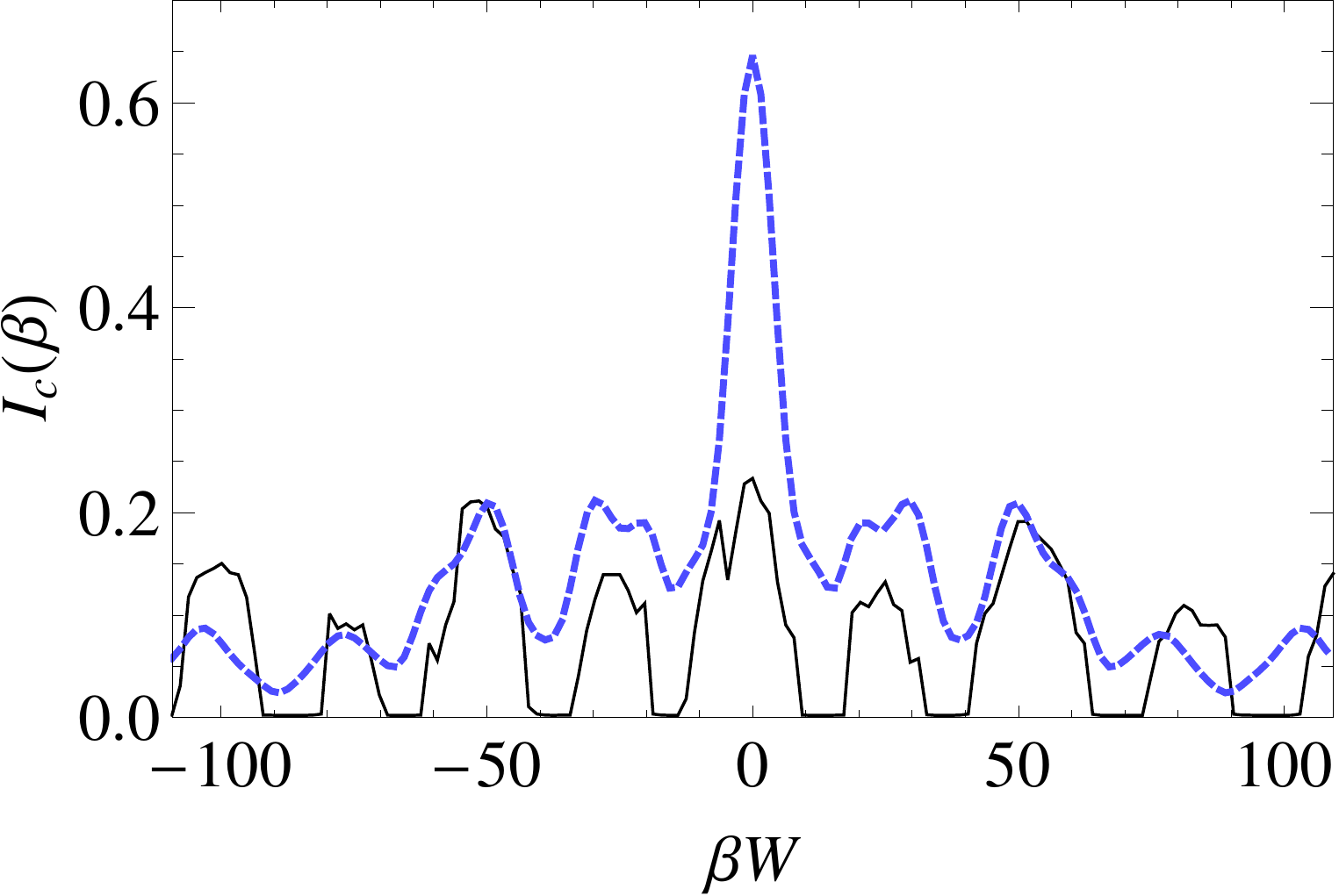}\caption{\label{fig:Kouwenhoven_consistency}Fraunhofer pattern obtained from
digitalization of the experimental results in figure 5(a) of Ref.
\onlinecite{Pribiag14_Edge_mode_SC_in_2DTI} (shown as a black line),
where the sample is believed to be in the SQH insulating regime. The
blue dashed line corresponds to the reconstructed Fraunhofer pattern
obtained from the reported current profile $J_{\text{DF}}\left(x\right)$
in figure 5.b in that reference.}
\end{figure}

In this section we present an analysis of the internal consistency
of the Dynes-Fulton method by comparing the original (i.e., experimental)
Fraunhofer pattern (obtained from a direct digitalization of the reported
experimental data) with the reconstructed pattern obtained from the
reported current profile $J_{\text{DF}}\left(x\right)$ (see Figs.
\ref{fig:Yacoby_consistency} and \ref{fig:Kouwenhoven_consistency}).
In Fig. \ref{fig:Yacoby_consistency} we present the data corresponding
to figure 2.c in Ref.~\onlinecite{Hart13_Induced_SC_at_QSH_edge}
as a continuous black line, whereas the dots correspond to the reconstructed
Fraunhofer pattern obtained from the reported current profile $J_{\text{DF}}\left(x\right)$
in figure 2.d in that reference, and Eq. \ref{eq:Fourier_transform}.

The corresponding analysis for the data reported by Pribiag \textit{et
al}\cite{Pribiag14_Edge_mode_SC_in_2DTI} is shown in Fig. \ref{fig:Kouwenhoven_consistency}.
The continuous black line is the digitized data in figure 5.a in that
reference, and the dot-dashed curve corresponds to the Fraunhofer
pattern reconstructed from the data reported in figure 5.b.

In general, we find that the data for the HgCdTe-HgTe system of Ref.~\onlinecite{Hart13_Induced_SC_at_QSH_edge}
shows more consistency with the Dynes-Fulton analysis than the data
of Ref.~\onlinecite{Pribiag14_Edge_mode_SC_in_2DTI} on the InAs-GaSb
system. For example, the extracted current distribution in Ref.~\onlinecite{Pribiag14_Edge_mode_SC_in_2DTI}
consistently spreads out way outside the nominal sample boundary which
is unphysical, and may be pointing toward the inapplicability of one
of the assumptions about the underlying physics.

\bibliographystyle{apsrev}

\end{document}